\def\beq{\begin{equation}}
\def\eeq{\end{equation}}
\def\bea{\begin{eqnarray}}
\def\eea{\end{eqnarray}}
\def\bq{\begin{quote}}
\def\eq{\end{quote}}

\def\gappeq{\mathrel{\rlap {\raise.5ex\hbox{$>$}}
{\lower.5ex\hbox{$\sim$}}}}

\def\lappeq{\mathrel{\rlap{\raise.5ex\hbox{$<$}}
{\lower.5ex\hbox{$\sim$}}}}

\def\bbz{fa Z \kern-8.9pt Z}

\documentstyle [12pt]{article}
\begin{document}
\thispagestyle{empty}
\begin{flushright}
{CERN-TH-98-161} \\
{May 1998} \\
\end{flushright}
\vspace{1cm}
\begin{center}
{\large Basis independent parametrisations } \\
\vspace{.2cm}
{\large of R parity violation in the soft SUSY breaking sector} \\
\end{center}
\vspace{1cm}
\begin{center}
{Sacha Davidson }\\
\vspace{.3cm}
{CERN Theory Division\\
CH-1211 Gen\`eve 23, Switzerland}
\end{center}
\hspace{3in}

\begin{abstract}
The magnitude of R-parity violating
coupling constants  depends on which
direction  in the space of weak doublets
with  hypercharge
$ = -2 $
corresponds to the Higgs.
To address this ``basis dependence'',
one can construct 
 combinations
of coupling constants that 
are invariant under these
basis transformations, and which parametrise
how much R parity violation is present in the Lagrangian
(analogous to Jarlskog invariants for CP
violation). This has previously been done
for the Higgs vev and the R parity
violating couplings constants  in the superpotential.
In this letter, I build invariants that
include soft SUSY breaking  interactions,
and briefly discuss their relation to invariants
involving the Higgs vev. This completes
the construction of invariants based on
the MSSM with baryon parity.
\end{abstract}


In the Standard Model, it is not possible to write
down any renormalisable interactions
that violate either baryon number ($B$)
or lepton number ($L$)\cite{Ross}. This is a consequence
of the gauge symmetries and the particle
content. In the supersymmetric
extension of the Standard Model \cite{SUSY} there
are many new particles, and it becomes possible
to have renormalisable $B$ or $L$
non-conserving  interactions. However,
since neither $B$ nor $L$ violation has
been observed in the laboratory, 
these interactions are often removed
by imposing a symmetry. 

There are various  symmetries that
prevent the $B$ or $L$ violating
renormalisable interactions in the
supersymmetric Standard Model \cite{IR}. 
The most common is a discrete symmetry called $R$-parity,
refered to as $R$  in this letter. 
It alots each particle  a multiplicative quantum
number :$(-1)^{2S + 3B +L}$, where $S$ is the 
particle spin. 

One can also allow the renormalisable
$B$ or $L$ violating interactions to
be present, but require the coupling constants
to be sufficiently small to statisfy
experimental bounds \cite{D}. In this case, it is not
desirable for the $B$ and $L$ violating
interactions to be simultaneously present, because
they mediate rapid proton decay. The bound
on the product  (lepton number violating
Yukawa-type coupling)$\times$(baryon number
violating Yukawa-type coupling) varies from
$10^{-10} $ to $10^{-25}$ \cite{Smirnov}, depending on
the generation indices of the coupling
constants. One therefore usually requires that
either $B$ or $L$ be conserved. In this
paper, I will assume that baryon number is
exactly conserved in the renormalisable interactions.

The superpotential for the Minimal Supersymmetric
Standard Model (MSSM) with $R$-parity
imposed is
\beq
W = \mu H_1 H_2 + h_u^{pq} H_2 Q_p U^c_q + h_d^{pq} H_1 Q_pD^c_q +
    h_e^{ij} H_1 L_i E^c_j ~~~. \label{1}
\eeq
The Lagrangian also contains kinetic terms,
gauge interactions, D-terms and soft SUSY breaking
terms of the form
\beq
{\rm soft~ masses} + B_H H_1 H_2 +  A_u^{pq} H_2 Q_p U^c_q + 
 A_d^{pq} H_1 Q_pD^c_q +
    A_e^{ij} H_1 L_i E^c_j ~~~. \label{2}
\eeq
I am abusively using capital letters for both
superfields (as in eqn \ref{1}) and scalar
component fields (as in eqn \ref{2}). Quark
generation indices are $p,q,r,s...$ and
lepton indices are $i,j,k...$. Whether indices
are up or down makes no difference. 

If instead of imposing $R$-parity, one
merely requires that baryon number be conserved, there
can be lepton number violating interactions
in the superpotential:
\beq
W_{L  \! \! \! \! /} = \epsilon^i H_2 L_i + \lambda^{ijk} L_i L_j E^c_k
                + \lambda^{'ipq} L_i Q_p D^c_q \label{3}
\eeq
and in the soft terms:
\beq
{\rm soft~ masses~ mixing}~ L^{\dagger}~ {\rm and}~ H_1 +  
 B^i H_2 L_i  + A_{\lambda}^{ijk} L_i L_j E^c_k
                + A_{\lambda'}^{'ipq} L_i Q_p D^c_q ~~~. \label{4}
\eeq

There are experimental upper bounds on these new couplings
from various processes \cite{D}, such as 
Flavour-Changing-Neutral-Currents (FCNC), 
lepton flavour violation and
lepton number non-conservation. However, some of these
coupling constants can be made zero by a basis choice,
so it is important to remember in which basis the bounds
apply.  In this letter I would like to approach this problem 
in a different way; I construct combinations of coupling constants
that are ``basis independent'' and that parametrise the amount
of $R$-parity violation present in the Lagrangian. They
are zero if $R$, or equivalently lepton number, is conserved.

A simple example of this approach is to take the
superpotential of equations (\ref{1}) and (\ref{3}) with
one quark and lepton generation. It appears
to have two $R$ violating interactions:
$\epsilon H_2 L$ and $\lambda' LQD^c$. 
(There is no  $\lambda^{ijk} L_iL_jE^c_k$ interaction because
$\lambda^{ijk}$  is antisymmetric on the $ij$ indices.) It is
well known that one of these can be rotated into
the other by mixing $H_1$ and $L$. If
\bea
H_1' = \frac{\mu}{\sqrt{\mu^2 + \epsilon^2}} H_1 +  
        \frac{\epsilon}{\sqrt{\mu^2 + \epsilon^2}} L \nonumber \\
L' = \frac{\epsilon}{\sqrt{\mu^2 + \epsilon^2}} H_1 -  \label{simple}
        \frac{\mu}{\sqrt{\mu^2 + \epsilon^2}} L ~~~,
\eea
then the Lagrangian expressed in terms of $H_1'$ and
$L'$ contains no $  H_2 L'$ term. One could instead dispose 
 of the $\lambda' L QD^c$ term.
The coupling constant
combination that is invariant under basis redefinitions
in $H_1,~L$ space, zero if $R$ parity is conserved,
and non-zero if it is not is $\mu \lambda' - h_d \epsilon =
(\epsilon, \mu) \wedge (\lambda', h_d)$.

``Basis-independent'' parametrisations of $R$ parity
violation have previously been constructed from subsets
of the parameters of the $R$ parity  non-conserving 
Minimal Supersymmetric Standard Model (MSSM). 
An invariant parmetrising
the R violation due to the misalignment between
the neutral vev and the superpotential $\mu$-term
 was constructed
in \cite{Banks} and subsequently much discussed \cite{Nardi}.
Invariants measuring the $R$ parity violation
between superpotential couplings, including
the one discussed in the previous paragraph,
were discussed in \cite{DE1,DE2}. The aim of this
letter is to construct the ``missing''
invariants involving soft terms. 

The invariants in \cite{Banks,DE1} measure
lepton number violation, whatever the lepton
flavour; the singlet lepton family index is summed.
 I will here follow the approach
of \cite{DE2} and build invariants that parametrise
lepton number violation in each family. Note that these
invariants do not measure lepton flavour violation
when lepton number is conserved.  
Invariants with lepton flavour indices are more
numerous, but have the advantage that their
relation to $R$ violating
coupling constants (which have lepton
flavour indices) is more direct.

In this letter I will first introduce
some notation, review the
geometric interpretation of the  invariants,
and then construct invariants
parametrising the $R$ parity
violation amoung the soft terms, and between
the soft terms and the superpotential terms.
Finally I will briefly discuss the relation
of the invariants introduced here to the one
of \cite{Banks}, and calculate cosmological bounds
on $R$ violating soft terms.

The lepton number violating interactions
in equations (\ref{3}) amd (\ref{4})
arise because the Higgs $H_1$ has the same
gauge quantum numbers as the doublet
sleptons $L_i$. If lepton number is {\em not}
a symmetry of the Lagrangian, then 
there is no longer any distinction based on
quantum numbers between the Higgs $H_1$  and a slepton $L_i$,
or between the higgsino and a doublet lepton.
One can therefore assemble the superfields $H_1$ and the $L_i$
in a four component vector
\beq
\phi^I = (L_1, L_2, L_3, H_1)~~~ ,~I:1..4
\eeq
and rewrite the superpotential as
\beq
W = \mu^I \phi_I H_2 + h_u^{pq} H_2 Q_p U^c_q + 
     \lambda_d^{Ipq} \phi_I Q_p D^c_q +
    \lambda_e^{IJk} \phi_I \phi_J E^c_k ~~~.  \label{1'}
\eeq
$\mu^I = ( \epsilon_i, \mu)$,  and $\lambda_d^{Ipq} = 
(\lambda^{'i pq}, h_d^{pq})$  are vectors in $\phi^I$
space. $\lambda_e^{IJk}$ is an anti-symmetric matrix,
with $\lambda_e^{4jk} = h_e^{jk}$, $\lambda_e^{ijk} =
\lambda^{ijk}$. 

The soft terms can be similarly rewritten as
\beq
\frac{1}{2}\phi^{I \dagger} m^2_{IJ} \phi^J  + B^I H_2 \phi_I 
 + A_u^{pq} H_2 Q_p U^c_q + 
     A_d^{Ipq} \phi_I Q_p D^c_q +
    A_e^{IJk} \phi_I \phi_J E^c_k + h.c.~~~,  \label{2'}
\eeq
where $\phi^I$ is now composed of scalar fields.
The reason for introducing this notation is that it makes
the geometrical significance of the couplings
constants clearer.  There is a four dimensional
vector space spanned by the hypercharge $ = -2 $
doublets $(H_1$ and $L_i)$. $\mu^I$ and $(\lambda_d^{pq})^I$
are vectors in this space, and they correspond to directions
that would like to be the Higgs---$i.e.$,
if one chooses a basis where the Higgs is
parralel to $\mu^I$, then the $R$ violating
masses $\mu^1, \mu^2,\mu^3$ are absent.

$[\lambda_e^k]^{IJ}$ is a little harder to visualise. It is
an antisymmetric $ 4\times 4$ matrix, and geometrically
corresponds to one or two planes which would
like to be spanned by  a Higgs and a lepton (see
\cite{DE2}  for a  discussion of
this). More practically, $[\lambda_e^k]^{IJ}$ is
a two index object (in $\phi$ space) that, when
contracted with a Higgs, becomes a vector
corresponding to a lepton. For instance, if
the Higgs direction is
\beq
\hat{H}^I \propto \mu^I \label{H}
\eeq
then  the lepton directions can be taken to be
\beq
(\hat{L}_k)^J \propto \mu_I [\lambda_e^k]^{IJ} \label{L}
\eeq
($\hat{L}_k$ with a hat is a basis vector, not neccessarily
of unit length. $L_k$ without a hat is a quantum number, or
sometimes a superfield or scalar field.)
$ [\lambda_e^k]^{IJ}$
is anti-symmetric, so $\mu^I$ is automatically perpendicular
in $\phi$ space  to $(\hat{L}_k)^J $.
If the singlet lepton basis is chosen such
that $ \hat{L}_k \cdot \hat{L}_m \propto \mu_I  [\lambda_e^k]^{IJ} 
[\lambda_e^m]^{JK*} \mu_K^* \propto
\delta^{km} $ (in the absence of $R$ violation, this would
be $h_e^{kj} h_e^{mj*} \propto \delta^{km}$), then
the lepton directions in $\phi$ space are also orthogonal. 
Note that transformations amoung the singlets $E_c^k$
rotate the $[\lambda_e^k]^{IJ}$ into each other
on their singlet index $k$.

\begin{figure}
\begin{picture}(400,300)(-150,-100)
\thicklines
\put(0,0){\vector(1,2){50}}
\put(0,0){\vector(0,3){150}}
\thinlines
\put(0,-50){\line(0,3){250}}
\put(0,0){\line(3,1){75}}
\put(0,0){\line(-3,-1){100}}
\put(-50,0){\line(3,0){250}}
\put(50,100){\line(0,-3){130}}
\put(50,-30){\line(3,1){90}}
\put(50,-30){\line(-3,0){135}}
\put(-35,120){$\mu$}
\put(20,80){$\lambda_d$}
\put(10,180){$H$}
\put(170,10){$L_1$}
\put(-120,-30){$L_2$}
\end{picture}
\caption{Non-orthogonality of $\mu^I$ and
$\lambda_d^{Ipq}$ in an $R$ violating model
with two lepton generations. 
The basis here is $ \hat{H} \propto \mu^I$, 
and  $ \hat{L}_k \propto \mu_I \lambda_e^{IJk}$.
The normalised projections of $\lambda_d^{Ipq}$ onto the
two directions identified as leptons 
parametrise the amount of $L_k$ ( or $R$) violation
and are independent of the basis chosen.}
\label{f1}
\end{figure}
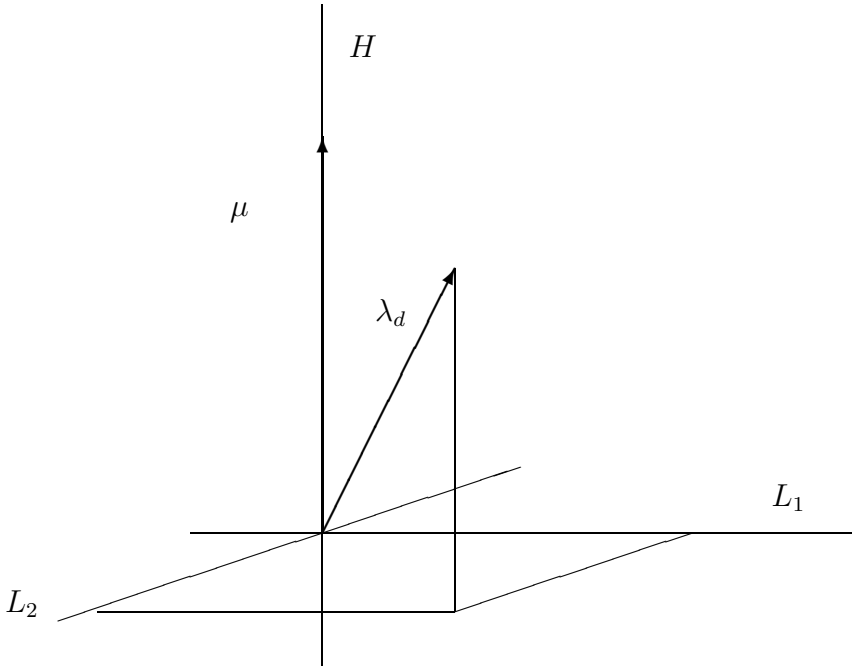

I have now constructed a geometrically motivated
orthogonal basis in $\phi$ space using a subset of coupling
constants from the Lagrangian. There is $R$
violation if this basis conflicts with the one
chosen by a different collection of coupling constants.
For instance, there are nine $(\lambda_d^{pq})^I$
vectors which could be chosen as the Higgs direction. 
If they have components in the directions
labelled as leptons by equation (\ref{L}), then $R$
is not conserved, and the scalar quantity that
parametrises $R$ violation in the $k$th lepton flavour is 
\beq
\sqrt{ \delta^{kpq}_{\mu \lambda_d}} =
\hat{L}_k \cdot \frac{\lambda_d^{pq}}{|\lambda_d^{pq}|} =
\frac{ \mu_I \lambda_e^{IJk} \lambda_d^{Jpq}}
 {|\mu||\lambda_e^k||\lambda_d^{pq}|} ~~~. \label{delta}
\eeq
See figure \ref{f1}.
The norms are defined in the obvious way; see
\cite{DE1}. The first reason for dividing by the
magnitude of $\lambda_d^{pq}$ is to avoid privileging
$\lambda_d^{pq}$ over $\mu$. Both  are equally good
choices for the Higgs direction, and this
invariant just parametrises the difference between 
the two. The second reason is to
give a normalised measure of how much $R$ violation
is present; (\ref{delta}) ranges from 0, when
$\mu$ and $\lambda_d^{pq}$ are parrallel, to
1 when they are orthogonal in the plane of
$\lambda_e$.

One can square the invariant (\ref{delta}), in which case the
coupling constant combination corresponds to a
closed supergraph; see \cite{DE1}. One can also
use two $\lambda_e^k$ matrices (geometrically planes)
to define a Higgs direction (the intersection of the
planes)and two leptons (the orthogonal directions within
the planes); in this case only the squared invariant
can be constructed, so it is arguably better to
chose the square of (\ref{delta}) as the invariant.
A more complete discussion of this geometry
and the invariants that can be constructed
from the superpotential can be found in
\cite{DE1,DE2}.

 There are more invariants than
$R$ violating coupling constants, because the
invariants are not all independent. To see
this, consider a model where there are (at least)
three different definitions of the Higgs:
$H$, $H'$ and $H''$. There are invariants
corresponding to the components of the vectors 
$H-H'$, $H' - H''$ and $H - H''$, but
since one of these vectors can be written
as the sum of the two others,  some of the
invariants can be expressed in terms of the others. 
The counting of independent invariants
constructed out of real coupling constants
is discussed in \cite{DE2}; there and
here, I neglect possible CP violating
phases. See \cite{Howie} for a counting
of the number of free parameters,
including phases,  in
the renormalisable Lagrangian of
the MSSM with baryon parity imposed. 

I would like to extend this geometric construction
to include the soft breaking terms. For the $B$ and
$A$ terms, this is straightforward, because 
one can build the same invariants as was done
with the $\mu$ and $\lambda$ terms in the superpotential
\cite{DE2}. There are many possibilities.
A minimal set, in terms of which the
rest can be expressed, and which is
complete in the sense that if all
the invariants are zero, then there
is no $R$ violation amoung the $A$ and
$B$ terms, could be the 27 invariants
that measure $L_{\ell}$ violation between $B$ and the $A_d$:
\beq
\tilde{\delta}_{BA_d}^{\ell pq} = \frac{|B^{I *} A_e^{IJ \ell  } 
   A_d^{Jpq *}|^2 }
{|B|^2  | A_e^{\ell}|^2 | A_d^{pq}|^2} \sim
\left| \hat{L}^{\ell} \cdot \frac{A_d^{pq}}{|A_d^{pq}|} \right|^2 
 \label{suggg1}
\eeq
and the nine invariants that parametrise $L_j$
violation in $A_e^{\ell}$:
 \beq
\tilde{\delta}_{BA_e}^{ jl} = \frac{B^{I *} A_e^{IJj} A_e^{JKl* } 
  A_e^{KLl} A_e^{LMj *} B^M
  - .5 B^I A_e^{IJ j*} A_e^{JK l} B^{K*} Tr[ A_e^j 
  A_e^{l *}] }
{|B|^2 Tr[A_e^{j*} A_e^j A_e^{l*} 
A_e^l]}  \sim
\left| \hat{L}^{\ell} \cdot \frac{A_e^{j}}{|A_e^{j}|} \right|^2 
\label{suggg2}
\eeq
where I am using a definition of lepton  based on the
soft terms :$\hat{L}^k \propto [A_e^k] \cdot B$. 
There are 36 independent invariants constructed out
of superpotential couplings, and the 36
invariants of similar form constructed out of $B$ and
$A$ terms listed above. To form a complete set,
one needs invariants asssociated with possible 
lepton number violating soft masses mixing
$H_1$ with the sleptons,  and one needs
invariants to measure $R$ violation
between the superpotential and the soft
breaking terms. 

Consider first these second invariants. In
principle it could be possible for the superpotential
couplings to make a unique choice of the Higgs, 
and for the $B$ and $A$ terms to do the same,
but for these two directions to be different.
It is sufficient to have three invariants,
for instance
\beq
\tilde{\delta}_{B \mu}^{\ell } = \frac{|B^{I *} \lambda_e^{IJ \ell } 
   \mu^{J *}|^2 }
{|B|^2  | \lambda_e^{\ell}|^2 | \mu|^2} \sim
\left| \hat{L}^{\ell} \cdot \frac{B}{|B|}  \right|^2  
 \label{super-soft}
\eeq
where $\ell:1..3$. In the basis of equations
(\ref{H}) and (\ref{L}), these  measure the projection
of $B$ along the three lepton directions. 

Finally we need invariants that parametrise
lepton number violation in the soft mass
matrix for the sleptons and Higgs $H_1$:
$\phi^{I \dagger} m^2_{IJ} \phi^J$. $m^2_{IJ}$  is
a hermitian matrix; the $R$ violating couplings
are the off-diagonal elements that mix $H_1$
with the sleptons. If I identify the Higgs
direction as $\mu^I$, and the leptons as
$(\hat{L}_k)^J \propto \mu_I \lambda_e^{IJk}$,
then the fractional amount of $R$ violation
in lepton family $k$ is
\beq
\tilde{\delta}_{\mu m^2}^k = 
\frac{|\mu_I \lambda_e^{IJk} m^2_{JK} \mu^K|^2}
{|\mu|^4 |\lambda_e^k|^2 |m^2|^2} \sim
\left| \hat{L}^{k} \cdot m^2 \cdot  \hat{H} \right|^2 \label{softmass}
\eeq
where $k:1..3.$

Note that these invariants measure lepton number
violation in each family (using
the definition of family of (\ref{L}), but not lepton flavour
violation where lepton number is conserved. 
The latter would be proportional to
$\hat{L}_k \cdot m^2 \cdot \hat{L}_j$.
 
Equations (\ref{softmass}), (\ref{super-soft}), (\ref{suggg1}),
and (\ref{suggg2}), plus equations (\ref{suggg1}) and
(\ref{suggg2}) with $B$ replaced by $\mu$ and
$A$ by $\lambda$ should form a complete set of real invariants
parametrising $R$ violation in the Lagrangian. There
are 78 of them. This is as expected; there are
naively 81 new coupling constants in equations
(\ref{3}) and (\ref{4}), and three of them
(commonly the $\epsilon_i H_2 L_i$) 
can be removed when choosing the Higgs direction
in $\phi$ space. Of course, all these
new coupling constants can have phases,
so the number of new parameters  is closer
to $2 \times 78$ (see \cite{Howie} for an
exact count). 

The list of invariants that I have constructed
does not include the wedge product of
$\mu$ and the vev of $\phi$, which was introduced in 
\cite{Banks}. If the direction in $\phi$
space that gets a vev $( \equiv v^I$) is
not parrallel to $\mu^I$, {\it i.e.}  $\mu \wedge v \neq 0$,
then the neutrino gets a mass (proportional
to $\mu \wedge v$). This  invariant
is phenomenologically relevant, so I would
like to check that it can be expressed in terms
of the invariants listed here.

The wedge product between $\mu$ and the vev measures
the sum of lepton number violation in all generations.
Since the other invariants I
have discussed measure $R$ violation in a specific
lepton generation, it is useful to have a flavour
dependent version of the invariant of \cite{Banks}:
\beq
 \frac{ \mu \cdot \lambda_e^{\ell} \cdot v}
 {|\mu| |\lambda_e^{\ell}| |v|} ~~~.
\eeq

Assuming that the neutral vev 
does not break  $R$ spontaneously,
it can only be misaligned with
$\mu^I$ if some coupling constants in
the scalar potential are misaligned with $\mu^I$.
$R$ violation between
$\mu^I$, $B^I$ and $m^2_{IJ} $ 
is parametrised by equations (\ref{super-soft}) and
(\ref{softmass}) so there should be some relation to
 $\mu \wedge v$. This is straightforward to solve
if only one sneutrino gets a vev.

The tree level potential for the vevs
of $H_2$ and $\phi^I$, respectively $\eta$ and
$v^I$,   is
\beq
V(v^I,\eta) =  \frac{1}{2} m_2^2 \; \eta^2 + B^I v_I \eta
+ \frac{1}{2} (m^2_{IJ} + \mu_I \mu_J) v^I v^J
+ \frac{g^2 + g^{'2}}{32}(\eta^2 - v^2)^2 \label{pot} ~~~.
\eeq
 This is minimised when
\beq
\left[  \frac{g^2 + g^{'2}}{8}(\eta^2 - v^2) 
- m_2^2 \right] \eta = - B_I v^I
\eeq
and
\beq
\left[  \frac{g^2 + g^{'2}}{8}(\eta^2 - v^2) 
- (m^2_{IJ} + \mu_I \mu_J) \right] v^J = \eta B^J ~~~. \label{not}
\eeq
Taking the inner product of (\ref{not}) with
$\mu \cdot \lambda_e^k$, dropping indices, and
assuming that the vev only overlaps with
the $k$th lepton generation, gives
\beq
\frac{\mu \cdot \lambda_e^k \cdot v}{|\mu| |\lambda_e^k | |v|}
 = \frac{\alpha \beta \pm \gamma
\sqrt{(\alpha^2 + \gamma^2) \rho -\beta^2}}
  {\alpha^2 +\gamma^2} ~~~~,
\eeq 
where
\beq
\alpha = \frac{g^2 + g^{'2}}{8}(\eta^2 - v^2) - 
\frac{\mu \cdot \lambda_e^{k} m^2 \lambda_e^k \cdot  \mu}
{|\lambda_e^k \cdot \mu|^2}~~~,
\eeq
\beq
\rho = \frac{ |\mu \cdot \lambda_e|^2}{|\mu|^2 |\lambda_e^k|^2} ~~~,
\eeq
and $\beta$ and $\gamma$ parametrise $R$ violation:
\beq
\beta^2 =  \frac{\eta^2}{|v|^2} \tilde{\delta}_{B \mu}^k
~B^2~~~,
\eeq
\beq
 \gamma^2 = \frac{|\mu |^2  | \lambda_e^{k}|^2 }
{ |\mu \cdot \lambda_e|^2} \tilde{\delta}_{\mu m^2}^k ~ |m^2|^2~~~\label{last}.
\eeq

The obvious question to ask, once these
invariants are
constructed is ``what are they good for?''.
In principle, they clarify how $R$ parity violation can be moved around
the Lagrangian. In practise, they
are messy and their relevance to phenomenology
is unclear  because $SU(2)$ is spontaneously broken;
the invariants  respect the
$SU(2) \times U(1)$ gauge symmetry of
the Lagrangian, but the propagating mass
eigenstates do not.

Invariants can be useful in calculating cosmological
bounds on $R$ violation \cite{CDEO}, because the
thermal mass eigenstate basis above the
electroweak phase transition is
not the same as the zero temperature one. One must therefore
take some care in identifying which interactions
violate $R$, or work in a basis independent formalism.
This is discussed in \cite{DE2} for the $L$ violating
superpotential couplings. Including soft breaking
interactions changes this analysis slightly, because
there are two additional mass terms  ($B$ and $m^2$)
that would like to choose a direction in $\phi$
space for the Higgs. I will briefly discuss
the modifications here. Bounds on $A$ terms
and trilinears were discussed in \cite{CDEO}.

The cosmological bounds on $L$ 
violation arise in models where the
observed baryon asymmetry was generated
before the electroweak phase transition. 
For the asymmetry to survive in the
presence of $B+L$ violating non-perturbative
electroweak effects, it must be an asymmetry
in at least one of the $B/3 - L_i$.  Therefore
interactions violating at least one of the
$B/3 - L_i$ must be out of equilibrium
just above the electroweak phase transition.
This gives a bound on the $B$ violating
trilinear $\lambda''$ couplings, and on the
lepton number violating couplings in one
generation.

The reason naive rate estimates can give
the wrong bound on $L$ violating couplings
can be understood in 
the one generation exactly supersymmetric
toy model where lepton number violation 
can be rotated between the $\epsilon LH_2$
term and the $\lambda' LQD^c$ term.
If one puts the
$L$ violation in the mass term, 
the rate for $L$ violation is
$\Gamma \sim \epsilon^2 /T$.
However if one rotates $\epsilon$ away,
one generates a a trilinear of order
$\lambda' \sim h_d \times \epsilon/\mu$,
for which the  rate should be of order
$\Gamma \sim \lambda^{'2} T$.
Requiring the first of these rates
to be out of equilibrium gives
$\epsilon < 10 $ keV, requiring the
second to be out of equilibrium gives
$\epsilon/\mu <10^{-5}$. For $\mu \sim 100$ GeV, 
these are not the same.  The point
is that the mass interaction is
stronger, so it determines what is
the Higgs, and lepton number violation
is in the trilinear, with the basis
independent magnitude $ y = \sqrt{\delta_{\mu \lambda_d} (h_d^2
+ \lambda^{'2})}$. $\delta_{\mu \lambda_d}$
is from equation (\ref{delta}) forone quark and
lepton generation. The $R$ violating
rate at temperatures above
 the electroweak phase transition is
therefore of order $ \Gamma_y \sim 10^{-2}
y^2 T$ where the $10^{-2}$ accounts for
various numerical factors \cite{CDEO}. The constraint
$\Gamma_y < H$, where $H \sim 10 T^2/m_{pl}$ is the expansion
rate of the Universe, gives
\beq
\sqrt{ \delta_{\mu \lambda_d} |\lambda_d|}< 10^{-7} ~~~. \label{bd1}
\eeq
For $|\lambda_d| \sim m_b/(\sqrt{2} \, v)$, this gives
$\delta_{\mu \lambda_d} < 10^{-11}$.

Now consider the case when there are
three possible mass terms that can determine
what is the Higgs: $\mu, B,$ and $m^2_{IJ}$.
 Above
the electroweak phase transition,
assuming $\mu^2 \sim B \sim m^2 \sim T^2$,
one can estimate an
interaction rate for the
$\mu_I \phi^I H_2$ from mass
corrections to a gauge boson-fermion-fermion
vertex. This gives
\beq
\Gamma_{\mu} \sim 10^{-2} \frac{|\mu|^2}{T}
\eeq
Similarly, one can estimate rates
for  $B$ and $m_{IJ}^2$ from the decay
of a scalar boson to two fermions to be of
order
\beq
\Gamma_B \sim 10^{-2} \frac{|B|^2}{T}, ~~~\Gamma_{m^2} \sim 10^{-2} \frac{m^2}{T}
\eeq
where $m^2$ is some eigenvalue of $m_{IJ}^2$, and the $B$
rate is in the basis where $B^I$ is the direction of
the Higgs.Each of these rates is
calculated as if the others
wer absent,  and none of them alone violates
lepton number. However, if they disagree on
what direction is the Higgs, then the lepton
number violating rates can be estimated as
in the toy model discussed above.

Suppose, for instance, that $\tilde{\delta}_{B \mu}^{\ell}$
(\ref{super-soft}) is non-zero, and $|B|$ is slightly
larger than $|\mu|^2$. Then $B$ determines what
is the Higgs, and the cosmological bound implies
$\tilde{\delta}_{B \mu}^{\ell} \Gamma_{\mu} < H$.
Since $\Gamma_{\mu} \sim \Gamma_B$, the bound from requiring
$ \tilde{\delta}_{B \mu}^{\ell} \Gamma_{B} < H$ is
approximately the same. A similar argument can be applied to
$\Gamma_B, \Gamma_{m^2}$ and $\tilde{\delta}_{B m^2}^{\ell}$
giving the bounds
\beq
\tilde{\delta}_{B \mu}^{\ell}, \tilde{\delta}_{B m^2}^{\ell}  < 10^{-14} 
\eeq
for one given lepton generation index $\ell$.
Notes that these bounds are more stringent than
(\ref{bd1}); the cosmological bounds
require that the mass terms be more aligned
with each other than with the trilinears.

%

In the $R$ violating MSSM, lepton number may
not be conserved, in which case there is
no distinction between the down-type
Higgs doublet $H_1$ and doublet leptons
$L_i$. These fields can be rotated into each 
other, changing the definition of
lepton number and the magnitude of $L$
violating coupling constants.  In \cite{DE1,DE2},
we constructed basis independent
combinations of coupling constants from the MSSM
superpotential that parametrised $L$ violation in
the renormalisable interactions. In this letter,
I have completed this process, constructing ``invariants''
that parametrise $R$ violation in the soft terms, 
and between the soft terms and the superpotential. 
A particular invariant is zero if there is
no lepton number violation among the
coupling constants used to construct it. 
If all the invariants are zero, there is
no $L$ violation in
the Lagrangian.  The invariants are
built by  using a minimal subset of
the Lagrangian to define
lepton number, and then seeing if
the remaining interactions conserve this
definition of $L_k$.


\begin{thebibliography}{222222}
\bibitem{Ross} for a discussion of B and L violating
       operators on the Standard Model and in supersymmetry, 
       see, $e.g.$ G.G.Ross, {\it Grand Unified Theories},
       Benjamin-Cummings, Menlo Park, CA,
       1985, chapters 7 and 11.
\bibitem{SUSY} for a review of the supersymmetric Standard Model,
see, $eg$ \\
 H.P. Nilles, {\it Phys. Rep.}  {\bf 110} (1984) 1; \\
 H.E. Haber and G.L. Kane, {\it Phys. Rep.} {\bf 117}(1985) 75; \\
 G.G. Ross, {\it Grand Unified Theories}, (Benjamin-Cummings, Menlo Park,
 CA, 1985); \\
 R. Barbieri, {\it Riv. Nuovo Cimento} {\bf 11} (1988)  1. 
\bibitem{IR} L. E. Ibanez, G. G. Ross, 
 {\it Nucl. Phys.} {\bf B368} (1992) 3.
\bibitem{D} H. Dreiner, in {\it Perspectives on Supersymmetry}, 
 Ed. by G.L. Kane, World Scientific, hep-ph/9707435;\\
G. Bhattacharyya, {\it  Nucl. Phys. Proc. Suppl.} {\bf 52A} (1997) 83, 
 hep-ph/9608415. 
\bibitem{Smirnov} A. Y. Smirnov, F. Vissani, 
{\it Phys. Lett.} {\bf B380} (1996) 317.  
\bibitem{Banks} T. Banks, Y. Grossman, E. Nardi, Y. Nir, 
{\it Phys. Rev.} {\bf D52} (1995) 5319, 1995; \\
H.-P. Nilles, N. Polonsky, {\it Nucl. Phys.} {\bf B484} (1997) 33. 
\bibitem{Nardi} see, {\it e.g.} \\
   B. de Carlos, P.L. White, 
{\it Phys. Rev.} {\bf D54} (1996) 3427; \\
R. Hempfling, {\it Nucl. Phys.} {\bf B478} (1996) 3; \\ 
E. Nardi, {\it Phys. Rev.} {\bf D55} (1997) 5772.
\bibitem{DE1}S. Davidson, J. Ellis, {\it Phys. Lett.}
{\bf B390}  (1997) 210.
\bibitem{DE2}S. Davidson, J. Ellis, {\it Phys. Rev.} {\bf D56} (1997) 4182.
\bibitem{CDEO}B.A. Campbell, S.Davidson, J. Ellis and K.A. Olive, 
   {\it Phys. Lett.} {\bf B 256} (1991) 457, 
   {\it Astroparticle Phys.} {\bf 1} (1992) 77,
    {\it Phys. Lett.} {\bf B297} (1992) 118 ; \\
  W. Fischler, G.F. Giudice, R.G. Leigh, S. Paban, 
  {\it  Phys. Lett.} {\bf B258} (1991) 45.\\ 
   H. Dreiner, G.G. Ross, {\it  Nucl. Phys.} {\bf B410}
   (1993) 188. 
\bibitem{Howie} H. Haber, work in progress.
\end{thebibliography}
\end{document}